\documentstyle[aps,12pt]{revtex}

\begin{document}
\draft
\author{Sergio De Filippo\cite{byline}}
\address{Dipartimento di Scienze Fisiche, Universit\`a di Salerno\\
Via Allende I-84081 Baronissi (SA) ITALY\\
and \\
Unit\`{a} INFM Salerno}
\date{\today}
\title{Quantum Computation using Decoherence-Free States of the Physical Operator
Algebra}
\maketitle

\begin{abstract}
The states of the physical algebra, namely the algebra generated by the
operators involved in encoding and processing qubits, are considered instead
of those of the whole system-algebra. If the physical algebra commutes with
the interaction Hamiltonian, and the system Hamiltonian is the sum of
arbitrary terms either commuting with or belonging to the physical algebra,
then its states are decoherence free. One of the considered examples shows
that, for a uniform collective coupling to the environment, the smallest
number of physical qubits encoding a decoherence free logical qubit is
reduced from four to three.
\end{abstract}


\section{Introduction}

Environment induced decoherence \cite{Zurek,Unruh,Landauer} is the main
obstruction to the physical viability of quantum computing \cite{Steane}. To
overcome this obstacle, quantum error correcting codes have been devised 
\cite{Shor,Steane}. Besides these {\it active} methods, where decoherence is
controlled by repeated application of error correction procedures, a more
recent {\it passive} approach has emerged, where logical qubits are encoded
in decoherence free (DF) subspaces \cite
{DuanGuo,ZanardiRasetti1,ZanardiRasetti2,Zanardi,LidarChuang,LidarBacon}. In
them coherence is protected by the peculiar structure of the coupling
Hamiltonian.

So far the notion of a DF state has been considered within the total Hilbert
space of the considered system, namely with reference to the whole operator
algebra of the system, whereas a more physical approach consists in
confining the consideration to the space of the states on the physical
algebra, that is the operator algebra involved in encoding and manipulating
qubits. The characterization of such state spaces corresponds to the
construction of the irreducible representations of the aforementioned
algebra. Quantum computing without active error correcting codes requires
the use of physical algebras admitting DF irreducible representations, which
therefore will be called DF algebras. The construction of such
representations is performed here by showing that suitable factorizations of
the total Hilbert space exist, where entanglement with the environment (or
equivalently decoherence, once this is traced out) is confined to only one \
factor, the other factor carrying an irreducible representation of the DF
algebra.

This more physical approach leads to a fruitful generalization of the notion
of a DF state. It is shown for instance that, for a generic uniform coupling
of an array of physical qubits to an arbitrary environment, while the
conventional notion of DF space requires at least four physical qubits to
encode a logical one \cite{Bacon}, three are enough in this new setting.

As to the plan of the paper, since it is addressed to a wide range of
theoreticians and experimentalists, the general notion of a quantum state as
a functional on a given $C^{\ast }$ algebra, instead of a density matrix on
a preassigned Hilbert space, is briefly introduced in the next section. This
is done with explicit reference to the ensuing relativity of the notion of
state purity, which is illustrated by the simplest possible example.

In the following section the concept of a decoherence free algebra is
presented with reference to a generic system, its Hamiltonian and its
coupling to the environment. In particular, the mentioned example and arrays
of qubits uniformly coupled to the environment are considered.

Then specific examples of three and four qubit arrays are analyzed, giving
explicit realizations of the DF algebras in terms of the original physical
qubit operators. In particular it is shown how the present generalized pure
states allow for the aforementioned DF logical qubit with only three
physical ones, while four physical qubits are shown to encode, in addition
to the known DF logical qubit\cite{Bacon}, a DF logical qutrit.

Finally some concluding remarks follow.

\section{C* Algebras and their pure states}

A quantum physical system is characterized by a $C^{\ast }$ algebra, namely
a normed complex associative algebra ${\cal A}$ with conjugation $\ast $ and
unity ${\bf 1}$, whose Hermitian elements are its observables, corresponding
in the usual operator setting to Hermitian bounded operators.\cite{Thirring}
Conjugation is an antilinear involution 
\begin{equation}
\ast :A\in {\cal A}\mapsto A^{\ast }\in {\cal A},{\cal \;(}A^{\ast })^{\ast
}=A,\;(cA+B)^{\ast }=\bar{c}A^{\ast }+B^{\ast }\;\forall c\in C,
\end{equation}
such that 
\begin{equation}
\left( AB\right) ^{\ast }=B^{\ast }A^{\ast },
\end{equation}
and the norm, endowing ${\cal A}$ with the structure of a Banach space, is
such that 
\begin{equation}
\left\| AB\right\| \leq \left\| A\right\| \left\| B\right\| ,\left\| A^{\ast
}\right\| =\left\| A\right\| ,\left\| AA^{\ast }\right\| =\left\| A\right\|
^{2},\left\| {\bf 1}\right\| =1.
\end{equation}

Boundedness is not a severe restriction, since every measurement apparatus
can detect only a finite range of values of an unbounded observable, by
which these observables play only a formal role as generators of groups of
unitary operators and can be eliminated altogether as primary physical
objects.

The states of the system correspondingly are positive and normalized linear
functionals: 
\begin{equation}
f:{\cal A}\rightarrow C,\;f(cA+B)=cf(A)+f(B)\;\forall c\in C,\;f\left(
AA^{\ast }\right) \geq 0,\;f\left( {\bf 1}\right) =1.
\end{equation}

States that can be written as linear convex combinations of different states 
\begin{equation}
f=\alpha g+(1-\alpha )h;\;f\neq g\neq h,\;0<\alpha <1
\end{equation}
are mixed states; otherwise they are pure.

Given a pure state $f$ one can uniquely construct, by the GNS procedure\cite
{Thirring}, a Hilbert space whose one dimensional projectors are pure states
(including the initial one), giving an irreducible representation of ${\cal A%
}$ in terms of bounded operators. This procedure is the $C^{\ast }$ algebra
counterpart of the Lie algebraic construction by raising and lowering
operators. The mentioned Hilbert space is identified with (the completion
of) the space of equivalence classes $\tilde{A}$\ of elements $A$ of ${\cal A%
}$, with respect to the equivalence relation 
\begin{equation}
\tilde{A}=\tilde{B}\Leftrightarrow f\left( \left[ A^{\ast }-B^{\ast }\right] %
\left[ A-B\right] \right) =0,
\end{equation}
\ where $\tilde{A}$ denotes the equivalence class of $A$, $A\tilde{B}\equiv $
$\widetilde{AB}$, the inner product is given by 
\begin{equation}
<\tilde{A}|\tilde{B}>\equiv f(A^{\ast }B)  \label{inner}
\end{equation}
and transition amplitudes by\cite{note} 
\begin{equation}
\left\langle \tilde{A}\right| C\left| \tilde{B}\right\rangle =f(A^{\ast }CB).
\label{transition}
\end{equation}

In general such an Hilbert space may not span the whole set of states,
namely inequivalent representations may ensue, starting from different
states. When this happens superselection rules are present, i.e. no
observable connects states belonging to inequivalent representations.

While superselection rules usually arise only in connection with infinitely
many degrees of freedom when ${\cal A}$ is defined in the usual way as
corresponding for instance to all possible measurements on a given set of
particles, this is not so if somehow it is restricted. In such a case the
restricted algebra, which is called here the physical algebra, may have
several different, (in general) reducible, representations inside the
Hilbert space corresponding to the unrestricted algebra.

The main idea in the present paper is to exploit the freedom in choosing the
physical algebra with reference to the notion of state pureness. A mixed
state of the whole algebra may be a pure state when restricted, as a
functional, to the physical algebra. In fact the pure states of the physical
algebra can be identified with equivalence classes of (in general not pure)
density matrices in the mentioned reducible representations. This may lead
to the physical equivalence between a non unitary evolution of the system
state in the usual sense (once the environment has been traced out) and a
unitary evolution with respect to the physical algebra if this one is
properly chosen.

To give the simplest possible illustration of the foregoing idea, consider a
two qubit system in the usual sense, namely a system consisting of two
atomic two-state systems. The corresponding operator algebra is generated by 
\begin{equation}
\sigma _{j}\otimes \sigma _{k},\;\;\;j,k=0,1,2,3,
\end{equation}
where $\sigma _{1},\sigma _{2},\sigma _{3}$ denote Pauli operators of a
single atomic system and $\sigma _{0}$ is the corresponding identity
operator. The usual product basis of the state space is given by 
\begin{equation}
\left| j,k\right\rangle \equiv \left| j\right\rangle \otimes \left|
k\right\rangle ,\;\;j,k=\pm 1  \label{product}
\end{equation}
where 
\begin{equation}
\sigma _{3}\left| j\right\rangle =j\left| j\right\rangle .
\end{equation}
On the other hand the operators 
\begin{eqnarray}
\pi _{1} &\equiv &1\otimes \sigma _{1},\;\pi _{2}\equiv \sigma _{3}\otimes
\sigma _{2},\;\pi _{3}\equiv \sigma _{3}\otimes \sigma _{3}  \nonumber \\
\tau _{1} &\equiv &\sigma _{2}\otimes \sigma _{1},\;\tau _{2}\equiv \sigma
_{3}\otimes 1,\;\tau _{3}\equiv \sigma _{1}\otimes \sigma _{1}  \label{pitau}
\end{eqnarray}
obey the same commutation relations 
\begin{equation}
\left[ \pi _{j},\pi _{k}\right] _{-}=2i\varepsilon _{jkl}\pi _{l},\;\left[
\tau _{j},\tau _{k}\right] _{-}=2i\varepsilon _{jkl}\tau _{l},\;\;\left[
\tau _{j},\pi _{k}\right] _{-}=0,
\end{equation}
as the single physical qubit operators 
\begin{eqnarray}
&&1\otimes \sigma _{1},\;1\otimes \sigma _{2},\;1\otimes \sigma _{3} 
\nonumber \\
&&\sigma _{1}\otimes 1,\;\sigma _{2}\otimes 1,\;\sigma _{3}\otimes 1.
\end{eqnarray}

Furthermore, since $\left[ \tau _{j},\tau _{k}\right] _{+}=\left[ \pi
_{j},\pi _{k}\right] _{+}=2\delta _{jk}$, the Casimir operators $\pi
_{1}^{2}+\pi _{2}^{2}+\pi _{3}^{2}$ and $\tau _{1}^{2}+\tau _{2}^{2}+\tau
_{3}^{2}$ assume the same value $3$ as the Casimir operator $\sigma
_{1}^{2}+\sigma _{2}^{2}+\sigma _{3}^{2}$ of the operator algebra
corresponding to a single traditional qubit. One can then identify the
operator algebra of the two qubit array with the direct product of the two
alternative $gl(2,C)$ algebras generated respectively by the $\pi $ and the $%
\tau $ operators.

Similarly the state space of the two qubit array can be realized as the
tensor product of two irreducible representations of the two alternative $%
gl(2,C)$ algebras, which can immediately be built by the GNS construction.

Consider for instance the state $f$ of the $\pi $ algebra uniquely defined
by 
\begin{equation}
f(\pi _{3})=-1,\;f(\pi _{1})=f(\pi _{2})=0.
\end{equation}
Then the equivalence classes of ${\bf 1}$ and $\pi _{+}=\left( \pi _{1}+i\pi
_{2}\right) /2$ give the usual basis 
\begin{equation}
\widetilde{{\bf 1}}{\bf =}\left| -1\right) ,\;\widetilde{\pi _{+}}=\left|
1\right) ,\;\pi _{3}\left| k\right) =k\left| k\right) ,
\end{equation}
as 
\begin{equation}
\left( 1\right| \pi _{3}\left| 1\right) =f(\pi _{-}\pi _{3}\pi _{+})=f\left( 
\frac{{\bf 1}-\pi _{3}}{2}\right) =1,\;\left( 1\right| \pi _{1,2}\left|
1\right) =f(\pi _{-}\pi _{1,2}\pi _{+})=0
\end{equation}
and of course 
\begin{equation}
\tilde{A}=\tilde{B}\Leftrightarrow A-B=c_{1}\pi _{-}+c_{2}\left( {\bf 1}+\pi
_{3}\right) ;\;c_{1},c_{2}\in C,\;\pi _{-}\equiv \pi _{+}^{\ast }.
\end{equation}

If the analogous notation is used for the $\tau $ algebra, one easily gets
the identification 
\begin{eqnarray}
\left| 1,1\right) &=&\frac{1}{\sqrt{2}}\left( \left| 1,1\right\rangle
+\left| -1,-1\right\rangle \right)  \nonumber \\
\left| 1,-1\right) &=&\frac{1}{\sqrt{2}}\left( \left| 1,1\right\rangle
-\left| -1,-1\right\rangle \right)  \nonumber \\
\left| -1,1\right) &=&\frac{1}{\sqrt{2}}\left( \left| 1,-1\right\rangle
+\left| -1,1\right\rangle \right)  \nonumber \\
\left| -1,-1\right) &=&\frac{1}{\sqrt{2}}\left( \left| 1,-1\right\rangle
-\left| -1,1\right\rangle \right) ,
\end{eqnarray}
where 
\begin{equation}
\left| j,k\right) \equiv \left| j\right) \otimes \left| k\right) ,\;\pi
_{3}\left| j,k\right) =j\left| j,k\right) ,\;\;\tau _{3}\left| j,k\right)
=k\left| j,k\right) .
\end{equation}

Assume now that the physical algebra is restricted to the one generated by
the $\pi $ operators. Then for instance the state 
\begin{equation}
\rho =\frac{\left| 1,1\right) \left( 1,1\right| +\left| 1,-1\right) \left(
1,-1\right| }{2}=\left| 1\right) \left( 1\right| \otimes \left( \frac{\left|
1\right) \left( 1\right| +\left| -1\right) \left( -1\right| }{2}\right)
\end{equation}
is trivially a pure state when restricted to the physical algebra, while its
expression in the original basis 
\begin{equation}
\rho =\frac{\left| 1,1\right\rangle \left\langle 1,1\right| +\left|
-1,-1\right\rangle \left\langle -1,-1\right| }{2}
\end{equation}
is entangled and then neither pure with respect to the first physical qubit
operator algebra, nor to the second one.

\section{Decoherence-free algebras}

Consider now the dynamics of a system $S$ coupled to a bath $B$, the {\it %
universe} evolving unitarily under the Hamiltonian $H=H_{S}{\bf \otimes 1}%
_{B}+{\bf 1}_{S}{\bf \otimes }H_{B}+H_{I}$, where $H_{S}${\bf \ }and $H_{B}$
denote respectively the system and the bath Hamiltonian, $H_{I}$ the
interaction Hamiltonian, ${\bf 1}_{S}$ and ${\bf 1}_{B}$ the identity
operators on the Hilbert space ${\cal H}_{S}$ of the system and ${\cal H}%
_{B} $ of the bath respectively. Let ${\cal A}_{S}\equiv gl({\cal H}_{S})$
denote the operator algebra of ${\cal H}_{S}$, (which for simplicity is
assumed to be finite dimensional) and ${\cal A}_{DF}$ the invariant
subalgebra of ${\cal A}_{S}$ consisting of operators commuting with $H_{I}$: 
\begin{equation}
\left[ {\cal A}_{DF},H_{I}\right] =0.
\end{equation}
As a subalgebra of ${\cal A}_{S}$, ${\cal A}_{DF}$ has a natural $C^{\ast }$
algebra structure, by which, if measurements on the system are confined to
those represented by operators in ${\cal A}_{DF}$, state spaces can be
identified with its irreducible representations.

(While the GNS construction gives a general procedure to construct the
representation of ${\cal A}_{DF}$ containing a given state of ${\cal A}_{DF}$
and, as described below, it is closely connected with what the
experimentalist is expected to do in the present context, representations of 
${\cal A}_{DF}$ in the final examples will be defined explicitly in terms of
physical qubit operators.)

A pure state of ${\cal A}_{DF}$, namely a state prepared by a complete set
of measurements of ${\cal A}_{DF},$ such remains under time evolution, if
the system Hamiltonian is the sum of an operator belonging to ${\cal A}_{DF}$%
, giving rise to unitary evolution, and an operator that commutes with $%
{\cal A}_{DF}$, which for such a state gives rise to no evolution at all.

As the simplest nontrivial example consider the above two qubit system,
assuming that 
\begin{eqnarray}
H_{S} &=&\sum_{j=1}^{3}\alpha _{j}\pi _{j}+\sum_{j=1}^{3}\beta _{j}\tau
_{j},\;\;\alpha _{j},\beta _{j}\in C,  \nonumber \\
H_{I} &=&\sum_{j=1}^{3}B_{j}\tau _{j},
\end{eqnarray}
where $\pi $ and $\tau $ operators are defined in Eq. (\ref{pitau}) and $%
B_{j}$ denote bath operators; in this case ${\cal A}_{DF}$ is the $gl(2,C)$
algebra generated by $\pi $ operators. Then, for a product state
\begin{equation}
\rho =\left| \psi \right) \left( \psi \right| \otimes \rho _{\tau },
\end{equation}
the interaction with the environment has no effect on the evolution of the
first factor, which then has a unitary evolution even though time evolution
of $\rho $, and specifically of $\rho _{\tau }$,  is not unitary. It should
be stressed that, while this appears to be rather trivial in terms of $\pi $
and $\tau $ operators, it is quite hidden \ if the state and the
Hamiltonians are expressed in terms of the original physical qubit operators 
$\sigma $.

In order to pass from an ad hoc example to a physically more relevant and
general setting, consider an array of $N$ qubits. Let $\sigma _{0}$, $\sigma
_{1}$, $\sigma _{2}$, $\sigma _{3}$ be defined as above. If these matrices
are intended to be, as usual, representations of pseudospin Hermitian
operators in the single qubit state space, the operator algebra for the
whole array is generated by 
\begin{equation}
M(i_{1},i_{2},...,i_{N})\doteq \bigotimes_{j=1}^{N}\sigma
_{i_{j}};\;\;\;i_{j}=0,1,2,3.
\end{equation}
Let 
\begin{equation}
S_{i}=\frac{1}{2}\sum_{j=1}^{N}M(i\delta _{1j},i\delta _{2j},...,i\delta
_{Nj})
\end{equation}
denote the total pseudospin, where $\delta _{jk}$ is the Kronecker symbol,
and assume, as frequently done in the literature \cite{ZanardiRasetti2}, a
uniform collective coupling to the environment 
\begin{equation}
H_{I}=\sum_{i=1}^{3}S_{i}B_{i},
\end{equation}
where the bath operators $B_{i}$ commute with ${\cal A}_{S}$ and then with $%
{\cal A}_{DF}$. As to the system Hamiltonian, under the usual hypothesis of
equivalent uncoupled qubits \cite{ZanardiRasetti2} 
\begin{equation}
H_{s}=\varepsilon S_{3},
\end{equation}
it commutes with ${\cal A}_{DF}$, which, as said above, avoids decoherence
of states of \ ${\cal A}_{DF}$, even with the possible addition of terms
belonging to ${\cal A}_{DF}$, like the scalar couplings 
\begin{equation}
\sum_{i=1}^{3}M(i_{1}=0,i_{2}=0,...,i_{j}=i,...,i_{k}=i,...i_{N}=0)
\end{equation}
due to the exchange interaction present in NMR computing \cite{NMR}. Let $%
{\cal A}_{E}$ denote the algebra generated by the {\it errors} $S_{i}$. Of
course ${\cal A}_{DF}\cap {\cal A}_{E}$ is generated by (the identity and
by) the Casimir operator 
\begin{equation}
S^{2}=\sum_{i=1}^{3}S_{i}^{2},
\end{equation}
by which, in order to factor the operator algebra as a product of such
subalgebras, the state space must be reduced to an $S^{2}$ eigenspace. To
this end the system Hilbert space ${\cal H}_{S}$, as the tensor product of $N
$ fundamental representations of $sl(2,C)$, can be decomposed as the
Clebsch-Gordan sum of irreducible representations of the algebra $sl(2,C)$
generated by the operators $S_{i}$: 
\begin{equation}
{\cal H}_{S}=\bigoplus_{j}\bigoplus_{k=1}^{n_{j}}{\cal D}_{j},
\label{CGordan}
\end{equation}
where the index $j$ fixes the eigenvalue of the Casimir operator: $S^{2}%
{\cal D}_{j}=j(j+1){\cal D}_{j}$.

The operator algebra of the generic eigenspace of $S^{2}$ can be identified
with the product of the representations of the DF and the error algebras on $%
\bigoplus_{k=1}^{n_{j}}{\cal D}_{j}$: 
\begin{equation}
_{j}{\cal A}_{S}\equiv gl\left( \bigoplus_{k=1}^{n_{j}}{\cal D}_{j}\right)
\sim \;_{j}{\cal A}_{DF}\otimes \;_{j}{\cal A}_{E}.  \label{algfact}
\end{equation}
In fact the $S^{2}$ eigenspace in its turn can be identified with the direct
product of an $n_{j}$ dimensional complex space and just one copy of the
irreducible representation 
\begin{equation}
\bigoplus_{k=1}^{n_{j}}{\cal D}_{j}\sim C^{n_{j}}\otimes {\cal D}_{j}
\label{spacefact}
\end{equation}
through the one to one correspondence $\left| k,m\right\rangle
\leftrightarrow \left| k\right\rangle \otimes \left| m\right\rangle $, where 
$\left| k,m\right\rangle $ denotes the eigenvector of $S_{3}$ with
eigenvalue $m$ in the $k$th copy of ${\cal D}_{j}$, while $\left|
m\right\rangle $ denotes the only such eigenvector in ${\cal D}_{j}$ and $%
\left| k\right\rangle $ is the $k$th element of a basis of $C^{n_{j}}$. To
be more precise, once the mutually orthogonal vectors $\left|
k,j\right\rangle $ are fixed, one defines $\left| k,m\right\rangle \equiv
(S_{-})^{j-m}$ $\left| k,j\right\rangle $ by means of the lowering operator $%
S_{-}=S_{1}-iS_{2}$.

Since the generic operator $O$ on $C^{n_{j}}$ gives through this
identification an operator $O\otimes {\bf 1}_{{\cal D}_{j}}$ on $%
\bigoplus_{k=1}^{n_{j}}{\cal D}_{j}$ commuting with ${\cal A}_{E}$, which is
generated by operators of the form ${\bf 1}_{C^{n_{j}}}\otimes Q$, and since
all operators can be realized in terms of the operators $%
M(i_{1},i_{2},...,i_{N})$, it follows that operators on $C^{n_{j}}$ can be
identified with (equivalence classes of) elements of \ ${\cal A}_{DF}$. This
proves that the generic $S^{2}$ eigenspace can be identified with the
product of two spaces, carrying irreducible representations of ${\cal A}%
_{DF} $ and ${\cal A}_{E}$ respectively. It should be stressed that coherent
superpositions of $S^{2}$ eigenstates with different eigenvalues do not
exist as states of ${\cal A}_{DF}$, as they live in different
representations.

As to the operational method to construct the $S^{2}=j(j+1)$ representation
of ${\cal A}_{DF}$, it is just the physical translation of the, only
seemingly formal, GNS procedure. To be specific, first an arbitrary (mixed
or pure) state of the chosen $S^{2}$ eigenspace has to be prepared. Then a
complete set of measurements corresponding to Hermitian elements of ${\cal A}%
_{DF}$ is performed in order to select a pure state of ${\cal A}_{DF}$.
Finally the whole representation is spanned by arbitrary unitary evolution
generated by Hamiltonian operators belonging to ${\cal A}_{DF}$. Of course
here unitarity is referred to ${\cal A}_{DF}$ only, since the coupling with
the environment is simultaneously producing, in general, a non unitary
evolution of the whole system-algebra ${\cal A}_{S}$\ and, to be more
specific, of the $S^{2}=j(j+1)$ representation of ${\cal A}_{E}$. Finally it
is worth to remark that possible terms in the system Hamiltonian belonging
to ${\cal A}_{E}$ give rise to a further unitary evolution in ${\cal A}_{S}$%
, which in the present context is physically irrelevant since it does not
affect ${\cal A}_{DF}$.

\section{Examples}

As a first example of a qubit array, collectively and uniformly coupled to
the environment, consider a system of three physical qubits. The
corresponding DF algebra is generated by 
\begin{eqnarray}
b_{23} &\doteq &4\vec{S}^{2}\cdot \vec{S}^{3}=\sum_{j=1}^{3}1\otimes \sigma
_{j}\otimes \sigma _{j},\;\;b_{31}\doteq 4\vec{S}^{3}\cdot \vec{S}%
^{1}=\sum_{j=1}^{3}\sigma _{j}\otimes 1\otimes \sigma _{j},\;\;  \nonumber \\
b_{12} &\doteq &4\vec{S}^{1}\cdot \vec{S}^{2}=\sum_{j=1}^{3}\sigma
_{j}\otimes \sigma _{j}\otimes 1,
\end{eqnarray}
where $\vec{S}^{j}$ denotes the pseudospin vector of the $j$th qubit, and
the Clebsch-Gordan decomposition in Eq. (\ref{CGordan}) reads 
\begin{equation}
{\cal H}_{S}={\cal D}_{3/2}\oplus {\cal D}_{1/2}\oplus {\cal D}_{1/2}={\cal H%
}_{3/2}\oplus {\cal H}_{1/2}.
\end{equation}

Since the factorization of $_{j}{\cal A}_{S}$ in Eq. (\ref{algfact}) is
trivial for $S^{2}=15/4$ ($j=3/2)$, as the error algebra generates the whole
operator algebra, the analysis is confined to the eigenspace ${\cal H}_{1/2}$
with $S^{2}=3/4$.

One can now apply the general GNS procedure. As a starting point take the
pure state of ${\cal A}_{DF}$ corresponding to an arbitrary normalized
vector of ${\cal H}_{1/2}$. The ensuing Hilbert space of equivalence classes
of elements of ${\cal A}_{DF}$, according to Section 2, gives the looked for
representation of ${\cal A}_{DF}$. (Of course even a density matrix on $%
{\cal H}_{1/2}$ that, as a state of ${\cal A}_{DF}$, is a pure state, can be
taken as an equivalent starting point.) While this procedure can be applied
in principle to much more general cases than the present qubit array, the
final result given below can easily be checked directly.\cite{note2}

Using the symbol $_{1/2}O$ for the representation of the generic operator $O$
in ${\cal H}_{1/2}$, for instance it can be checked that, if one defines the
invariant operator 
\begin{equation}
E_{123}\doteq \sum_{i,j,k=1}^{3}\varepsilon _{ijk}\sigma _{i}\otimes \sigma
_{j}\otimes \sigma _{k},
\end{equation}
with $\varepsilon _{ijk}$ denoting the usual completely antisymmetric
symbol, the ${\cal H}_{1/2}$ representations of \ invariant operators 
\[
_{1/2}\tau _{1}=(\;_{1/2}b_{12}-\;_{1/2}b_{23})/\sqrt{12}
\]
\[
_{1/2}\tau _{2}=\;_{1/2}E_{123}/\sqrt{12}
\]
\begin{equation}
_{1/2}\tau _{3}=(\;_{1/2}b_{23}-2\;_{1/2}b_{31}+\;_{1/2}b_{12})/6
\label{tau}
\end{equation}
are the generators of an $su(2)$ algebra, 
\begin{equation}
\left[ \;_{1/2}\tau _{i},\;_{1/2}\tau _{j}\right] =2i\sum_{k=1}^{3}%
\varepsilon _{ijk\;}\;_{1/2}\tau _{k},  \label{commrel}
\end{equation}
with the Casimir given by 
\begin{equation}
_{1/2}\tau ^{2}\equiv \sum_{j=1}^{3}\;_{1/2}\tau _{j}^{2}=3{\bf \hat{1}}.
\label{Casimir}
\end{equation}
The corresponding universal enveloping algebra ${\cal A}\left( _{1/2}{\bf %
\tau }\right) $, which coincides with $\;_{1/2}{\cal A}_{DF}$, is then the
operator algebra of a two state system and the total operator algebra $_{1/2}%
{\cal A}$ is given by the product of this algebra and the universal
enveloping algebra ${\cal A}\left( _{1/2}{\bf S}\right) $ of the total
pseudospin algebra: 
\begin{equation}
_{1/2}{\cal A}={\cal A}\left( _{1/2}{\bf \tau }\right) \otimes {\cal A}%
\left( _{1/2}{\bf S}\right) =\;_{1/2}{\cal A}_{DF}\otimes {\cal A}\left(
_{1/2}{\bf S}\right) ,
\end{equation}
as a particular instance of Eq. (\ref{algfact}). As a consequence the state
space ${\cal H}_{1/2}$ can be identified with the tensor product of two
two-dimensional representation spaces ${\cal H}_{1/2}({\bf \tau })$ and $%
{\cal H}_{1/2}({\bf S})$ respectively of ${\cal A}\left( _{1/2}{\bf \tau }%
\right) $ and ${\cal A}\left( _{1/2}{\bf S}\right) $: 
\begin{equation}
{\cal H}_{1/2}={\cal H}_{1/2}({\bf \tau })\otimes {\cal H}_{1/2}\left( {\bf S%
}\right) ,
\end{equation}
which coincides with Eq. (\ref{spacefact}) for $j=1/2$ and $n_{j}=2$.
According to what has been illustrated above, this factorization has far
reaching physical consequences: if all measurement processes are limited to
(Hermitian) elements of ${\cal A}_{DF}$, then a state $\rho =\left| \psi
\right\rangle \left\langle \psi \right| \otimes \rho _{{\bf S}}$, which is
the product of a pure state in ${\cal H}_{1/2}({\bf \tau })$ and an
arbitrary density matrix in ${\cal H}_{1/2}\left( {\bf S}\right) $, is a
pure state of the physical algebra ${\cal A}_{DF}$. If in particular the
initial state has this structure (possibly with $\rho _{{\bf S}}$ being
itself a pure state of \ ${\cal A}\left( _{1/2}{\bf S}\right) $, this
corresponding to an arbitrary pure state in ${\cal H}_{1/2}$), then, in
spite of the decoherence of $\rho _{{\bf S}}$ (or equivalently the
entanglement with the environment if this is not traced out) produced by the
coupling of the environment to the pseudospin operators, the state maintains
phase coherence as to the physical algebra, which is then DF. This means
that the considered three qubit array encodes a DF logical qubit, compared
to the four qubits needed within the conventional approach \cite{Bacon}.

As a further example, consider now a four qubit array, whose Clebsch-Gordan
decomposition is 
\begin{equation}
{\cal H}_{S}={\cal D}_{2}\oplus {\cal D}_{1}\oplus {\cal D}_{1}\oplus {\cal D%
}_{1}\oplus {\cal D}_{0}\oplus {\cal D}_{0}={\cal H}_{2}\oplus {\cal H}%
_{1}\oplus {\cal H}_{0}.
\end{equation}
In this case, while the factorization is trivial and useless for the $%
S^{2}=6 $ ($j=2$) representation, it is still trivial but fruitful for the
carrier space ${\cal H}_{0}$ of the two degenerate $S^{2}=0$
representations, where it gives rise to the DF states already considered in
the literature. To be specific it can be checked that the ${\cal H}_{0}$
representations of invariant operators 
\[
_{0}\tau _{1}\doteq (\;_{0}b_{14}+\;_{0}b_{23}-\;_{0}b_{12}-\;_{0}b_{34})/(4%
\sqrt{3}), 
\]
\[
_{0}\tau _{2}\doteq
(\;_{0}E_{234}+\;_{0}E_{124}-\;_{0}E_{134}-\;_{0}E_{123})/(8\sqrt{3}), 
\]
\begin{equation}
_{0}\tau _{3}\doteq -(\;_{0}b_{14}+\;_{0}b_{12}+\;_{0}b_{13})/3
\end{equation}
obey the same relations as their analogues in Eq.s (\ref{commrel},\ref
{Casimir}), whose enveloping algebra once again is the operator algebra of a
DF logical qubit. As represented in ${\cal H}_{0}$ the DF subalgebra
coincides with the total operator algebra, the representation of the total
pseudospin algebra being the trivial (scalar) one.

For the four qubit array, apart from the reproduction of a DF qubit of
vanishing pseudospin, the present approach gives also rise to a DF qutrit.
Consider in fact the 9-dimensional $S^{2}=2$ ($j=1$) eigenspace ${\cal H}%
_{1} $ containing three degenerate 3-dimensional representations. It can be
checked, for instance, that the ${\cal H}_{1}$ representation of invariant
operators 
\[
_{1}\tau _{1}\doteq \;_{1}E_{134}/(-2\sqrt{3}), 
\]
\[
_{1}\tau _{2}\doteq (\;_{1}E_{134}-3\;_{1}E_{124})/(4\sqrt{6}), 
\]
\begin{equation}
_{1}\tau _{3}\doteq (\;_{1}E_{234}+\;_{1}E_{123})/(4\sqrt{2})
\end{equation}
obey the usual commutation rules of $su(2)$ generators as in Eq. (\ref
{commrel}), while $_{1}\tau ^{2}\equiv \sum_{j=1}^{3}\tau _{j}^{2}=8{\bf 
\hat{1}}$. In this case the 9-dimensional state space $_{1}{\cal H}$ \ can
be identified with the product of the 3-dimensional irreducible
representations of the DF algebra and the total pseudospin algebra. In
perfect analogy to what said for the three qubit array one can arrange in
the considered $S^{2}=2$ eigenspace a DF qutrit, namely a tridimensional
state space of the DF algebra. Of course in this case the whole
representation algebra $_{1}{\cal A}_{DF}$ cannot be produced by linear
combination of the $sl(2,C)$ generators (and the identity) only, but
products of two of them must be included too.

\section{Conclusion}

In conclusion what has been shown can be of use both with reference to the
considered examples and more generally as a method to identify for given
systems several alternative DF spaces, which can give rise to more chances
for finding physically viable realizations of quantum computing. In
particular the possibility to test DF qubit encoding in arrays of just three
physical qubits may represent a substantial bonus in the near future.

More generally a new viewpoint about decoherence is advocated and shown to
be effective. It is shown that the very notion of decoherence should be
defined in more physical terms starting from the notion of physical algebra.
Before asking if a state of a given system is pure or not we should
preliminarily fix the operator algebra with respect to which we are defining
the state. The main result of the paper is that if pureness is not defined
in an abstract setting, starting from the operator algebra of the whole
universe, but on the contrary from the operator algebra generated by the
actual measurements that the experimentalist is going to perform, a
thoroughly new and promising perspective appears. This result is relevant
not only with reference to quantum computing but even to the foundations of
quantum mechanics and the analysis of open quantum systems in general. In
particular the approach in terms of representations of \ DF algebras may
shed some light on the physical relevance of quantum coherence, which in
principle, due to the structure of the Hamiltonian, could be present in
unexpected situations if system algebras can be factored as the product of
uncoupled collective algebras, one of them decoupled from the environment
too.

Acknowledgments.{\LARGE -- }I thank L. Viola for pointing out to my
attention Ref.\cite{Viola}, which is relevant to the subject treated in this
paper and, due to my ignorance, was omitted from the original references.

The present paper was supported by M.U.R.S.T., Italy and I.N.F.M., Salerno.


\begin{references}
\bibitem{note}  Neither Eq. (\ref{inner}), nor Eq. (\ref{transition}) depend
on the choice of specific representatives $A\in \tilde{A}$, $B\in \tilde{B}$%
. In fact $f\left( \left[ E^{\ast }+\bar{\lambda}F^{\ast }\right] \left[
E+\lambda F\right] \right) \geq 0\;\forall \lambda \in C;\;E,F\in {\cal A}%
\Rightarrow \left| f\left( F^{\ast }E\right) \right| ^{2}\leq f\left(
E^{\ast }E\right) f\left( F^{\ast }F\right) $, by which that independence
follows if $E$ and $F$ are meant to be differences between elements of $%
\tilde{B}$ and $\tilde{A}$ respectively.
\bibitem{note2}  One can easily check that for the three qubit array, in the
product basis 
\[
\left| \psi \right\rangle \equiv \frac{1}{\sqrt{2}}\left( \left|
-1,-1,1\right\rangle -\left| 1,-1,-1\right\rangle \right) \Rightarrow
S^{2}\;\left| \psi \right\rangle =3\left| \psi \right\rangle ,\;\tau
_{3}\left| \psi \right\rangle =-\left| \psi \right\rangle ,\frac{\tau
_{1}-i\tau _{2}}{2}\;\left| \psi \right\rangle =0, 
\]
where $\tau $ operators are defined as in Eq.s (\ref{tau}) without
restriction to the $j=1/2$ subspace. By which in this case one can
explicitly construct the representation by the raising operator, namely in
terms of the basis 
\[
\left| \psi \right\rangle ,\;\frac{\tau _{1}-i\tau _{2}}{2}\left| \psi
\right\rangle =\frac{1}{\sqrt{6}}\left( 2\left| -1,1,-1\right\rangle -\left|
-1,-1,1\right\rangle -\left| 1,-1,-1\right\rangle \right) 
\]
the matrix representation of the operators $\tau _{1}$, $\tau _{2}$, $\tau
_{3}$ is given by the Pauli matrices.
\end{references}
\end{document}